\documentclass{gen-p-l}
\usepackage[dvips]{graphics}
\newtheorem{theorem}{Theorem}[section]

\theoremstyle{definition}

\theoremstyle{remark}

\numberwithin{equation}{section}

\newcommand{\Ir}{\mathbb{Z}}
\newcommand{\Cx}{\mathbb{C}}

\newcommand{\HH}{\mathcal{H}}
\newcommand{\KK}{\mathcal{K}}

\newcommand{\Proj}{{\rm Proj}}
\newcommand{\Span}{{\rm span}}
\newcommand{\abs}[1]{\lvert#1\rvert}
\newcommand{\floor}[1]{\left\lfloor{#1}\right\rfloor}
\newcommand{\ceil}[1]{\left\lceil{#1}\right\rceil}
\newcommand{\ket}[1]{\lvert#1\rangle}
\newenvironment{alphlist}{%
  \begin{enumerate}%
}{%
  \end{enumerate}}

\begin{document}
\title{Interfaces and droplets in quantum lattice models}
\author{Bruno Nachtergaele}
\address{Department of Mathematics, University of California, Davis,
Davis, CA 95616-8633}
\email{bxn@math.ucdavis.edu}
\thanks{This material is based on work supported by
the National Science Foundation under Grant No. DMS0070774.
\newline\indent
Copyright \copyright\ 2000 Bruno Nachtergaele. Reproduction of this 
article in its entirety, by any means, is permitted for 
non-commercial purposes.}
\subjclass{82B10, 82B24, 82D40}
\date{August 16, 2000.}
\begin{abstract}
This paper is a short review of recent results on interface states
in the Falicov-Kimball model and the ferromagnetic XXZ Heisenberg model.
More specifically, we discuss the following topics:
1) The existence of interfaces in quantum lattice models that can
be considered as perturbations of classical models. 2)
The rigidity of the 111 interface in the three-dimensional
Falicov-Kimball model at sufficiently low temperatures. 3)
The low-lying excitations and the scaling of the gap 
in the 111 interface ground state in the ferromagnetic XXZ 
Heisenberg model in three dimensions. 4) The existence  
of droplet states in the XXZ chain and their properties.
\end{abstract}

\maketitle

\section*{Introduction}

Domain walls or interfaces play an important role in many transport 
phenomena and are essential for the understanding of dynamical processes
such as magnetic ordering and magnetization reversal. In many
interesting situations quantum mechanics is important to correctly 
describe the stability, fluctuations, and excitations of magnetic
interfaces,  and we expect that this is even more true for a reliable
description of the  dynamics. Equilibrium states describing interfaces in
the Ising model are  well-known since the seminal work by Dobrushin
\cite{Dob}, and many interesting results on interfaces in classical models
have been proved  since then. In comparison, there have been very few
results for quantum  models. In the last few years however, some progress
has been made and  it is the purpose of this paper to review the main
results that have been obtained.

In the next four sections we will discuss the following results: 
\begin{enumerate}
\item 
Existence of interfaces in quantum lattice models that can
be considered as perturbations of classical models. 
\item Rigidity of the 111 interface in the Falicov-Kimball model.
\item Gapless excitations of the 111 interface
in the XXZ Heisenberg model.
\item Droplet states in the XXZ chain.
\end{enumerate}

\section{Perturbations of classical models}

Often, quantum lattice models can be thought of as perturbations of a
classical spin system when some parameter in the Hamiltonian is not too
large (or too small). Under certain conditions its behavior can then 
be described by a suitable perturbation theory. The following result is due
to Borgs, Chayes, and Fr\"ohlich \cite{BCF}. In order to simplify the
presentation we do not describe the result in its most general form here.
Let $n$ be a fixed integer $\geq 2$. For every finite subset 
$\Lambda\subset\Ir^d$, let the Hilbert space for the system in volume 
$\Lambda$ be given by
$$
\HH_\Lambda = \bigotimes_{x\in \Lambda}\HH_x,\quad \HH_x = \Cx^n,
\mbox{ for all } x .
$$
Consider translation invariant local Hamiltonians of the form
\begin{equation}
H_\Lambda=\sum_{x,y\in\Lambda,\abs{x-y}=1}h_{xy}
 + \lambda \sum_{A\subset\Lambda}V_A
\label{BCFmodel}\end{equation}
where $h_{xy}$ is a translation invariant isotropic nearest
neighbor interaction that is diagonal in a tensor product basis labeled
by ``classical configurations'', i.e., there is a function $h(\sigma_0,
\sigma_1)$, such that 
$h_{xy}\ket{\sigma}=h(\sigma_x,\sigma_y)\ket{\sigma}$. Here,
$\sigma=\{\sigma_x\}_{x\in\Lambda}, 1\leq\sigma_x\leq n$, and  
$\ket{\sigma}=\bigotimes_{x\in\Lambda}\ket{\sigma_x}\in\HH_\Lambda$, 
and $\{\ket{1},\ldots,\ket{n}\}$ is a basis of $\Cx^n$. Assume 
that this classical nearest neighbor interaction has exactly two ground state,
of the form $\ket{\sigma_0}\otimes\ket{\sigma_0}$. The quantum perturbation
given by the $V_A$ is also assumed to be translation invariant
and sufficiently short range in the sense that, for some $r\geq 0$, 
sufficiently large,
$$
\sum_{A\subset\Ir^d,\, 0\in A} \Vert V_A\Vert e^{r D(A)}<+\infty
$$
where $\Vert\,\cdot\,\Vert$ is the usual operator norm, and $D(A)$ is the 
size of the smallest tree subgraph of $\Ir^d$
containing the vertices of $A$. 

\begin{theorem}[Borgs-Chayes-Fr\"ohlich  \cite{BCF}]\label{thm:BCF}
Under the assumptions described above, and
if $d\geq 3$, the model defined by (\ref{BCFmodel}) has Gibbs states with
a rigid interface in the principle coordinate directions, provided
that the inverse temperature $\beta$ is sufficiently large and 
$\beta\lambda$ is sufficiently small.
\end{theorem}                                                      

The proof of this theorem relies on recently developed techniques 
to derive Pirogov-Sinai type results for quantum lattice models
\cite{DFF, BKU}.

The next two results are examples of interesting situations where
perturbation theory as described above, does not apply. They also
illustrate two features that distinguish quantum models from
their classical counterparts: 1) in a  system that has classically several
equivalent energy minimizing configurations, ground state selection will
typically occur in the quantum model in the same way as tunneling in a
symmetric double well potential leads to a unique ground state, and 2)
finite-state models can have continuous symmetries and gapless energy
spectra.

\section{Rigidity of the 111 interface in the Falicov-Kimball model}

It is expected that the three-dimensional Ising model does not have Gibbs
states with an interface in the 111 direction, due to a large ground state
degeneracy for the obvious boundary conditions that would favor such an
interface. A proof of the Solid-on-Solid version of this statement was given
by Kenyon \cite{Keny}. In contrast, in the Falicov-Kimball and Heisenberg
XXZ models, in  dimensions $\geq 2$, this degeneracy is lifted by the
``quantum'' terms, i.e., ground state selection occurs. Our next result is
that, for the half-filled, neutral Falicov-Kimball model in three dimensions
the 111 interface is rigid at sufficiently low temperatures \cite{DMN}.

The Falicov-Kimball model consists of two subsystems living on the
same lattice $\Ir^d$: ``ions'', described by Ising type variables 
$W(x)\in\{0,1\}$, and ``electrons'', described by fermion operators 
$c^+_x,c_x$. The ions and electrons interact via an on-site Coulomb 
term as in the  Hubbard model. The local Hamiltonians are
\begin{equation}
H_\Lambda=-\sum_{\{x,y\}\subset\Lambda,\vert x-y\vert =1}
c^+_x c_y
+2U\sum_{x\in\Lambda}W(x)c^+_x c_x
\label{FKmodel}\end{equation}
We only consider the half-filled neutral case, i.e., the number of ions and
the number of electrons are both taken to be equal to half the number of
lattice sites and we take $U>0$. Kennedy and Lieb proved that, for $d\geq
2$, this model has an Ising type phase transition, with two low-temperature
phases dominated by the two checkerboard configurations for the ions
\cite{KL}.

In order to study the 111 interface in the three-dimensional model, we 
integrate out the fermion degrees of freedom using the methods of 
Messager and Miracle-Sol\'e \cite{MM}, and obtain an effective 
Hamiltonian for the ions in terms of the the Ising variables
$s_x=(-1)^{\vert x\vert}(2W(x)-1)$:
$$
H^{\rm eff}_\Lambda(\{s_x\})=
 -J(U,\beta)\sum_{\{x,y\}\cap\Lambda\neq\emptyset,\vert x-y\vert=1}s_xs_y
+\sum_{B\cap\Lambda\neq\emptyset,\vert B\vert \geq 2}
R_B(U)
$$
where, and $0<J(U,\beta)=1/(4U) +$ terms of higher order in $U^{-1}$ 
starting with $U^{-3}$ and also including a weak $\beta$ dependence.
$R_B$ depends on $\{s_x\mid x\in B\}$, is translation covariant and satisfies
$$
\sum_{B\ni 0}\vert R_B\vert e^{r n(B)} < \infty
$$
for some $r>0$, and where $n(B)$ is the length of the shortest closed path
in the lattice visiting all sites in $B$ at least once, and where $R_B$
satisfies a bound of the form
$$
\abs{R_B(U)} \leq \mbox{constant}\times\left(\frac{\rm constant}{U}
\right)^{n(B)-1}.
$$
Analysis of the terms with $n(B)=4$, indicates that ground state selection 
indeed occurs for the model with 111 boundary conditions. The following
theorem shows that the 111 interface is stable against thermal fluctuations
at sufficiently low temperatures.

\begin{theorem}[Datta-Messager-Nachtergaele \cite{DMN}] 
There exist constants $U_0$, $\! C$, $c$, and $D$, such that for all
$U> U_0$, and $\beta> DU^3$, there exists a Gibbs state $\omega_{111}$
of the half-filled neutral three-dimensional Falicov-Kimball model 
defined by (\ref{FKmodel}) with the following properties:
\begin{eqnarray*}
\omega_{111}(s_x)\geq \phantom{-}1-Ce^{-c\beta/U^3}&\mbox{if}&
x_1+x_2+x_3\geq 1\\
\omega_{111}(s_x)\leq -1+Ce^{-c\beta/U^3}&\mbox{if}&
x_1+x_2+x_3\leq -1
\end{eqnarray*}
\end{theorem}

The Falicov-Kimball model also exhibits phase separation, i.e., 
spontaneous formation of domains of different phases,
in the ground state for certain intervals of densities
\cite{Kend, KH}.

\section{Gapless excitations of the 111 interface in the XXZ Heisenberg 
model}

The spin 1/2 XXZ Heisenberg model defined by
\begin{equation}
H_\Lambda=-\sum_{x,y\in\Lambda,\vert x-y\vert =1}
\frac{1}{\Delta}(S_x^{1} S_y^{1} + S_x^{2} S_y^{2})+ S_x^{3} S_y^{3}
\label{XXZmodel}\end{equation}
with $\Delta > 1$, can also be regarded as a perturbation of the Ising
model. Kennedy proved that this model has a phase transition for all $d\geq
2$, $\Delta >1$ \cite{Kend}. Theorem \ref{thm:BCF} applies and interface
states in the principal coordinate directions exist for $d\geq 3$, and
$\beta$ and $\beta/\Delta$ large enough. One can also show that for $d\geq
2$, ground state selection occurs for suitable conditions that favor a
diagonal (11, 111, etc.) interface. The model differs from the
Falicov-Kimball model, however, in one important aspect: the existence of a
diagonal interface is accompanied by the breaking of a continuous symmetry
(rotations about the third axis), and consequently has a gapless spectrum
above the ground states. These gapless excitations were first described by
Koma and Nachtergaele in the two-dimensional case \cite{KN2, KN3}. Matsui
proved their existence in all dimensions $\geq 2$ \cite{Mat}.

I conjecture that the XXZ model has a 111 interface Gibbs state in three
dimensions at low temperatures. This cannot be proved using the standard
expansion techniques, which all rely on the existence of a gap. Therefore,
it is reasonable to first try to understand the gapless excitations of the
model. 

The third component of the total spin or, equivalently, the number of down
spins, is a conserved quantity of the XXZ model. Therefore, it is natural to
consider the interface problem and the excitation spectrum in the canonical
ensemble where this quantity is fixed. The following theorem holds
in the canonical ensemble. An equivalence-of-ensembles result is an
essential part of its proof.

\begin{theorem}[Bolina-Contucci-Nachtergaele-Starr \cite{BCNS1}]
Let $0<q<1$ be such that $\Delta=(q+q^{-1})/2$. Then the XXZ 
model with Hamiltonian (\ref{XXZmodel}), possesses
excitations localized in a cylinder with axis in the 111 direction
and with a circular intersection of radius
$R$ with the interface with the energy gap $\gamma_R$ bounded by
$$
\gamma_R\leq 100\frac{q^{2(1-\delta(q,\nu))}}{(1-q^2)}\frac{1}{R^2},      
\quad \textrm{for}\quad R>70,
$$
where $\delta(q,\nu)$ is an exponent between $0$ and $1/2$, depending on
the filling factor $\nu$ of the interface plane and the parameter $q
\in(0,1)$.
\end{theorem}

See \cite{BCN, BCNS1, BCNS2} for more details and related results.

\section{Droplet states in the XXZ chain}

The last result we would like to present is concerned with droplet 
states rather than interfaces. One does not expect that a localized
droplet in a translation invariant system can be described by a stationary 
state, i.e., an eigenvector, of the Hamiltonian. On the other hand, it
seems obvious that for a ferromagnetic model such as the XXZ Heisenberg
Hamiltonian, any state of a given finite energy in a sufficiently
large volume should be a superposition of states that predominantly 
consist of a small number of domains each described by one of the ground 
states. We have obtained an explicit result of this kind in the 
one-dimensional case.

Let $H^{++}_{[1,L]}$ be the following Hamiltonian with ``+''
boundary fields for a chain of $L$ spins:
\begin{equation}
H^{++}_{[1,L]}=-\sum_{x=1}^L
\frac{1}{\Delta}(S_x^{1} S_{x+1}^{1} + S_x^{2} S_{x+1}^{2})+ 
(S_x^{3} S_{x+1}^{3} - \frac{1}{4}) -A(\Delta)(S^3_1+S^3_L),
\label{dropletmodel}\end{equation}
where $S^i_x$ are the standard spin-1/2 matrices and $A(\Delta)=
\frac{1}{2}\sqrt{1-\Delta^{-2}}$. $H^{++}_{[1,L]}$ commutes with
$\sum_{x=1}^LS_n^3$, and therefore the subspaces $\HH_{L,n}$ of
$\HH_{[1,L]}$ consisting of all states with a fixed number, $n$, of down
spins, are invariant. The ground state in each of these invariant subspaces
is a delocalized droplet of $n$ down spins, i.e., this ground state is
dominated by configurations where all $n$ down spins are in a subinterval of
length $n + O(1)$, but all such intervals are roughly equally probable.
Thus, we expect that the droplet of size $n$ behaves as a particle freely
moving in an interval of length $L-n+1$. Our results on the spectrum show
that this particle has mass of order $q^{-n}$, where $q\in(0,1)$ is related
to the anisotropy by $\Delta=(q+q^{-1})/2$. To state the precise result, let
$\lambda_{L,n}(1) \leq \lambda_{L,n}(2) \leq \dots $ denote the eigenvalues
of $H^{++}_{[1,L]}$ restricted to $\HH_{L,n}$, and let $\psi^{++}_{L,n}(1),
\psi^{++}_{L,n}(2), \dots $ be the corresponding eigenvectors. It is
convenient, though not necessary, to describe the spectral projection
corresponding to the $L-n+1$ lowest eigenvalues rather than the individual
eigenvectors. Define
$$
 \HH^{k}_{L,n} = \Span \{ \psi^{++}_{L,n}(j) : 1 \leq j \leq k\}\, .
$$
and denote by $\Proj(\HH^{k}_{L,n})$ the corresponding orthogonal
projection. We have no exact expressions for the this spectral projection
for finite $n$, but we can describe it with an error of no more than order
$q^{n/2}$ in the operator norm. To state the result we need to introduce the
``droplet'' subspace of $\HH_{L,n}$. We start with the definition of the
kink and antikink states, which are ground state of the model with a
different choice of boundary fields \cite{ASW, GW}. For every interval
$[a,b]\subset\Ir$, and $n=0,1, \ldots, b-a+1$, the kink and antikink states,
$\psi^{+-}_{[a,b]}(n)$ and $\psi^{-+}_{[a,b]}(n)$ respectively, are given by
the following expressions:
\begin{eqnarray*}
\psi^{+-}_{[a,b]}(n) = 
  \sum_{a \leq x_1 < \dots < x_n \leq b} q^{\sum_{k=1}^n (b+1-x_k)}
  \left( \prod_{k=1}^n S_{x_k}^- \right) \ket{\uparrow \dots 
  \uparrow}_{[a,b]} \\
\psi^{-+}_{[a,b]}(n) = 
  \sum_{a \leq x_1 < \dots < x_n \leq b} q^{\sum_{k=1}^n (x_k+1-a)}
  \left( \prod_{k=1}^n S_{x_k}^- \right) \ket{\uparrow \dots \uparrow}_{[a,b]} 
\end{eqnarray*}
We now define the ``droplet'' states, describing a localized droplet
of size $n$ centered at $x$: for $n\geq 0$ and $\floor{n/2} \leq x 
\leq L - \ceil{n/2}$ define
\begin{equation}
 \xi_{L,n}(x) \
   = \psi^{+-}_{[1,x]}(\floor{n/2})
   \otimes \psi^{-+}_{[x+1,L]}(\ceil{n/2})\, .
 \end{equation}
Here, for any real number $x$, $\floor{x}$ is the greatest integer $\leq x$,
and $\ceil{x}$ is the least integer $\geq x$. The droplet subspace is now
defined as the span of the droplet states at all possible positions:
\begin{equation}
\KK_{L,n} = \Span \{ \xi_{L,n}(x) : \, 
\floor{n/2}\leq x \leq L-\ceil{n/2} \}\, .
\label{KLn}\end{equation}
Note that $\dim \KK_{L,n} = L-n+1$, which is the number of subintervals 
of length $n$ of an interval of length $L$. The orthogonal projection
onto $\KK_{L,n} $ is denoted by $\Proj(\KK_{L,n})$.

\begin{theorem}[Nachtergaele-Starr \cite{NS}] \hfill\vspace{-1mm}
\begin{alphlist} 
\item
The first $L-n+1$ eigenvalues of $H^{++}_{[1,L]}$ restricted to $\HH_{L,n}$
are all approximately equal to $A(\Delta)$. More precisely
we have the following:
$$
\lambda_{L,n}(1),\dots, \lambda_{L,n}(L-n+1) 
  \in  [A(\Delta) - O(q^n),A(\Delta) + O(q^n)]\, .
$$
\item
For sufficiently large $n$, the first $L-n+1$ eigenvalues are separated 
from the rest of spectrum by a gap of approximate magnitude
$\gamma=1-\Delta^{-1}$:
$$
\lambda_{L,n}(L-n+2) \geq A(\Delta) + \gamma - O(n^{-1/4}).
$$
\item
The spectral subspace $\HH^{L-n+1}_{L,n}$ spanned by the first 
$L-n+1$ eigenvectors, for sufficiently large $n$, is close to the space 
$\KK_{L,n}$ defined in (\ref{KLn}). More precisely we have the 
the following bound on the norm of the difference of the corresponding
projections:
$$
\|\Proj(\KK_{L,n}) - \Proj(\HH^{L-n+1}_{L,n})\|
= O(q^{n/2})\, . 
$$
\end{alphlist}
\end{theorem}

The gap of approximately $\gamma=1-\Delta^{-1}$ shown in part (b) of the
Theorem is the energy required to overturn one spin in either the interior
or the exterior of the droplet \cite{KN1, KN4}. This energy is less than the
energy required to create an additional droplet at infinite distance, which
is $2A(\Delta)$, i.e., droplets attract.

Part (c) of the Theorem means that the eigenvectors can be approximated by a
superposition of the states $\xi_{L,n}(x)$, which in turn are given by an
explicit formula, with increasing accuracy as $n\to\infty$. Note that the
Theorem also implies that, for any sequence of states with energies
converging to $A(\Delta)$, we must have that the distances of these states
to the subspaces $\KK_{L,n}$ converges to zero, i.e., in the limit the
states contain  exactly one droplet, which may either be localized or not.

\begin{figure}[tb]
\resizebox{12truecm}{8truecm}{\includegraphics{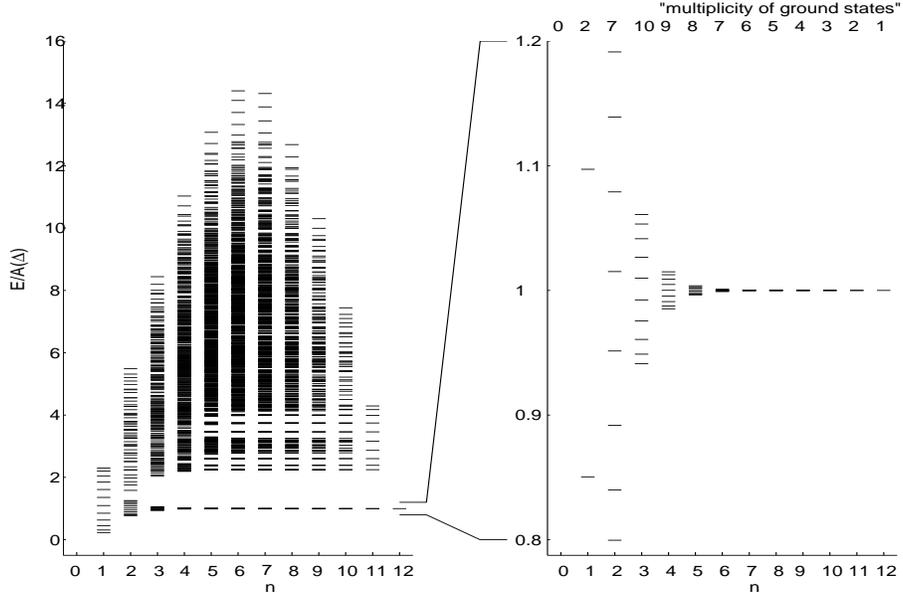}}
\parbox{11truecm}{\caption{\baselineskip=5 pt\small
\label{fig:dropspec}
Spectrum for $H^{++}_{[1,12]}$ when $\Delta = 2.125$ ($q=1/4$).
The (trivial) $n=0$ state is not shown. Its energy is $-A(\Delta)$.}
}
\end{figure}

Figure \ref{fig:dropspec} illustrates these features of the spectrum 
for $L=12$ and $\Delta=2.125$. The multiplet of droplet states is clearly
visible, as well as the gap above it. 

\bibliographystyle{amsalpha}

\begin{thebibliography}{99}

\bibitem{ASW} F.C. Alcaraz, S.R. Salinas, and W.F. Wreszinski, 
\textit{Anisotropic ferromagnetic quantum domains}, 
Phys. Rev, Lett. {\bf 75} (1995), 930--933. 

\bibitem{BCN} O. Bolina, P. Contucci, and B. Nachtergaele, \textit{Path 
integral representation for interface states of the 
anisotropic Heisenberg model},
to appear in  Rev. Math. Phys., \texttt{arXiv:math-ph/9908004}.

\bibitem{BCNS1} O. Bolina, P. Contucci, B. Nachtergaele, and S. Starr,
\textit{Finite-volume excitations of the 111 interface in the quantum 
XXZ model}, Comm. Math. Phys. {\bf 212} (2000), 63--91, 
\texttt{arXiv:math-ph/9908018}.

\bibitem{BCNS2} O. Bolina, P. Contucci, B. Nachtergaele, and S. Starr,
\textit{A continuum approximation for the excitations of the 
$(1,1,\ldots,1)$ interface in the quantum Heisenberg model},
Electronic Journal of Differential Equations, Conf. 04, (2000), 1-10, 
\texttt{arXiv:math-ph/9909018}.

\bibitem{BCF}
C. Borgs, J. Chayes, and J. Fr\"ohlich,
\textit{Dobrushin states in quantum lattice systems},
Commun. Math. Phys. \textbf{189} (1997), 591--619.

\bibitem{BKU} 
C. Borgs, R. Koteck\'y, and D. Ueltschi,
\textit{Low temperature phase diagrams of quantum perturbations
of classical spin systems},
Commun. Math. Phys., \textbf{181} (1996) 409--446.

\bibitem{DFF}
N.~Datta, R.~Fern\'andez, and J.~Fr{\"o}hlich,
\textit{Low-temperature phase diagrams of quantum lattice systems. {I}.
Stability for quantum perturbations of classical systems with finitely-many 
ground states}, J. Stat. Phys. {\bf 84} (1996), 455--534.

\bibitem{DMN} N. Datta, A. Messager, and
B. Nachtergaele, \textit{Rigidity of the 111 interface in the 
Falicov-Kimball model},
J. Stat. Phys., \textbf{99} (2000), 461-555, 
\texttt{arXiv:math-ph/9804008}.

\bibitem{Dob} 
R.L. Dobrushin, 
\textit{Gibbs state describing the coexistence of phases for a 
three--dimensional Ising model},
Theor. Prob. Appl. {\bf 17} (1972), 582.

\bibitem{GW} 
C.-T. Gottstein, R.F. Werner,
\textit{Ground states of the infinite q-deformed Heisenberg ferromagnet},
\texttt{arXiv:cond-mat/9501123}.

\bibitem{Kend}
T.~Kennedy,
\textit{Phase separation in the neutral Falicov-Kimball model},
J. Stat. Phys. \textbf{91} (1998), 829--843,
\texttt{arXiv:cond-mat/9705315}.

\bibitem{KH}
T. Kennedy and K. Haller,
\textit{Periodic Ground States in the Neutral Falicov-Kimball Model in 
Two Dimensions}, submitted to J. Stat. Phys.,
\texttt{arXiv:cond-mat/0004104}. 
    
\bibitem{KL}
T.~Kennedy and E.~H. Lieb,
\textit{An itinerant electron model with crystalline or magnetic long-range
order}, Physica \textbf{138A} (1986), 320--358.

\bibitem{Keny}
R. Kenyon,
\textit{Local statistics of lattice dimers},
Ann. Inst. H. Poincar\'e, Probab. Statist., 
\textbf{33} (1997), 591--618.  

\bibitem{KN1} T. Koma and B. Nachtergaele,
\textit{The spectral gap of the ferromagnetic XXZ chain}, 
Lett. Math. Phys. \textbf{40} (1997), 1--16.

\bibitem{KN2} T. Koma and B. Nachtergaele,
\textit{Low-lying spectrum of quantum interfaces},
Abstracts of the AMS, \textbf{17} (1996), 146, 
and unpublished notes.

\bibitem{KN3} T. Koma and B. Nachtergaele,
\textit{Interface states of quantum lattice models}, In Matsui, T. (eds.) 
\textit{Recent Trends in Infinite
Dimensional Non-Commutative Analysis}. RIMS Kokyuroku \# 1035, Kyoto,
1998, 133--144.

\bibitem{KN4} T. Koma and B. Nachtergaele,
\textit{The complete set of ground states of the ferromagnetic XXZ chains},
Adv. Theor. Math. Phys., \textbf{2} (1998), 533--558,
\texttt{arXiv:cond-mat/9709208}

\bibitem{Mat} 
T. Matsui, 
\textit{On the spectra of the kink for ferromagnetic $XXZ$ models},
Lett. Math. Phys. {\bf 42} (1997), 229--239.

\bibitem{MM}
A. Messager and S. Miracle-Sol\'e,
\textit{Low temperature states in the Falikov-Kimbal model},
Rev. Math. Phys. {\bf{8}} (1996) 271--299.

\bibitem{NS} B. Nachtergaele and S. Starr,
\textit{Droplet states of the XXZ Heisenberg chain},  
\texttt{arXiv:math-ph/0009002}.

\end{thebibliography}

\end{document}